\documentstyle[12pt]{article}
\begin{document}
\pagestyle{plain}
\setcounter{page}{1}
\newcounter{bean}
\baselineskip16pt
\begin{titlepage}

\begin{flushright}
PUPT-1648\\
hep-th/9609076
\end{flushright}
\vspace{20 mm}

\begin{center}
{\huge Four-dimensional greybody factors }

\vspace{5mm}
{\huge and the effective string}

\end{center}

\vspace{10 mm}

\begin{center}
{\large Steven S.~Gubser\footnote{e-mail: {\tt ssgubser@puhep1.princeton.edu}} 
and Igor R.~Klebanov\footnote{e-mail: {\tt klebanov@puhep1.princeton.edu}}
}

\vspace{3mm}

Joseph Henry Laboratories\\
Princeton University\\
Princeton, New Jersey 08544

\end{center}

\vspace{2cm}

\begin{center}
{\large Abstract}
\end{center}

\noindent
Recently Maldacena and Strominger found that the calculation
of greybody factors for $D=5$ black holes carrying three U(1) charges
gives striking new evidence for their description as multiply
wound effective strings. Here we show that a similar result
holds for $D=4$ black holes with four $U(1)$ charges.
In this case the effective string may be thought of as the triple
intersection of the 5-branes in M-theory compactified on $T^7$.

\vspace{2cm}
\begin{flushleft}
September 1996
\end{flushleft}
\end{titlepage}
\newpage
\renewcommand{\baselinestretch}{1.1} 

\renewcommand{\epsilon}{\varepsilon}
\def\fixit#1{}
\def\comment#1{}
\def\equno#1{(\ref{#1})}
\def\equnos#1{(#1)}
\def\sectno#1{section~\ref{#1}}
\def\figno#1{Fig.~(\ref{#1})}
\def\D#1#2{{\partial #1 \over \partial #2}}
\def\df#1#2{{\displaystyle{#1 \over #2}}}
\def\tf#1#2{{\textstyle{#1 \over #2}}}
\def\d{{\rm d}}
\def\e{{\rm e}}
\def\i{{\rm i}}
\def\Leff{L_{\rm eff}}

\section{Introduction}
\label{Intro}

Recently remarkable progress has been achieved in modeling
the Hawking radiation within the string theory or M-theory context.
There exist supersymmetric $D=5$ black holes 
with non-vanishing horizon area
which may be embedded into string theory
using intersecting D-branes \cite{sv}. Their low-energy dynamics
is described by small fluctuations of a long intersection string
\cite{cm,ms}. 
In
\cite{cm} it was shown how Hawking emission from the near-extremal
black holes takes place in the stringy description.  The model
involves $n_1$ 1-branes marginally bound to $n_5$ 5-branes, with some
longitudinal momentum along the 1-branes carried by left moving
open strings. In the non-extremal case, right movers are also
present, so that a left moving and a right moving open string may
collide to produce an outgoing closed string \cite{cm,hk}. 
This mechanism leads to a thermal
distribution for the massless outgoing particles \cite{cm}, as expected of 
Hawking radiation. The inverse of this process, which gives the
leading order contribution to the absorption of closed strings, was
also found to be in agreement with the semiclassical gravity, up to
an overall normalization \cite{dmw}.  More recently, 
Das and Mathur \cite{dm} carefully normalized the leading
emission and absorption rates, both in semiclassical gravity and
in the D-brane picture, and found perfect agreement!  The specific picture
used in \cite{dm} follows that suggested in \cite{dm1,ms}:  the low-energy
dynamics of the D-brane configuration is captured by a single string with
winding number $n_1 n_5$ which is free to vibrate only within the 5-brane
hyperplane.  

The calculation of Das and Mathur was carried out in the near-extremal
regime,

\begin{equation}
r_0 \ll r_n \ll r_1, r_5
\end{equation}

\noindent
where $r_0$ is the radius of the horizon, and $r_n, r_1, r_5$ are
three other radii determined by the charges. Very recently, 
Maldacena and Strominger \cite{mast} generalized the calculation to
a less restrictive choice,

\begin{equation}
r_0, r_n \ll r_1, r_5 \ .
\end{equation}

\noindent
The large difference in the scales is necessary to suppress the
antibranes \cite{hms}. 
Dropping the restriction $r_0\ll r_n$ allows the left and
right moving temperatures of the effective string to be comparable.  The
calculation of greybody factors in this regime reveals a dependence on
$T_L$ and $T_R$ which provides striking new evidence in favor of the
effective string model of $D=5$ black holes.

The purpose of this letter is to present similar evidence
for supersymmetric $D=4$ black holes with regular horizons
\cite{cy,ct}.
Such black holes may be embedded into
string theory (for a treatment of the non-extremal case, see
\cite{lowe}).
In \cite{at,kt,gkt} it was argued, however, that it is
advantageous to view these $D=4$ black holes as dimensionally reduced
configurations of intersecting branes in M-theory.  A specific
configuration useful for explaining the Bekenstein-Hawking entropy is
the $5\bot 5\bot 5$ intersection \cite{kt}: there are $n_1$ 5-branes
in the $(12345)$ hyperplane, $n_2$ 5-branes in the $(12367)$
hyperplane, and $n_3$ 5-branes in the $(14567)$ hyperplane. One also
introduces a left moving momentum along the intersection
string (in the $\hat 1$ direction). If the length of this direction is
$L_1$, then the momentum is quantized as $2\pi n_K/L_1$, 
so that $n_K$ plays the role of the fourth $U(1)$
charge.  Upon compactification on $T^7$ the metric of the $5\bot 5\bot
5$ configuration
reduces to that of the $D=4$ black hole with four charges.  Just like
in the D-brane description of the $D=5$ black hole, the low-energy
excitations are signals propagating along the intersection string. In
M-theory the relevant states are likely to be small 2-branes with
three holes glued into the three different hyperplanes \cite{kt}. As a
result, the effective length of the intersection string is
$\Leff= n_1 n_2 n_3 L_1$. 
This fact, together with the assumption that these
modes carry central charge $c=6$, is enough to reproduce the
extremal 
Bekenstein-Hawking entropy, $S=2\pi\sqrt{n_1 n_2 n_3 n_K}$ \cite{kt}.
In our previous paper \cite{us} we showed that this ``multiply-wound
string'' model of the four-charge $D=4$ black hole  correctly
reproduces the Hawking radiation of both neutral and Kaluza-Klein
charged scalars. The calculation of \cite{us} was performed in
the near-extremal regime,
$r_0 \ll r_4 \ll r_1, r_2, r_3$, where $r_0$ is the
horizon radius while $r_1, \ldots, r_4$ are four other radii
related to the charges.
Here we generalize it to a less restrictive choice of parameters:

\begin{equation}
r_0, r_4 \ll r_1, r_2, r_3 \ . 
\label{newchoice}\end{equation}

\noindent
As in the work of \cite{mast},
this relaxes the condition $T_R \ll T_L$.  The dependence of the greybody
factors on $T_L$ and $T_R$ is characteristic of the effective string model.

\section{The semiclassical gravity analysis}

We start with the 11-dimensional
configuration of three sets of 5-branes intersecting along a common
1-brane and carrying momentum along it. The 
non-extremal metric of this configuration was constructed in 
\cite{ct1}.\footnote{Our notation differs from that
in \cite{ct1} by the replacements
${\cal P}_i\rightarrow r_i$, $\tilde {\cal Q}\rightarrow r_4$ and
$\mu\rightarrow r_0$.}
Compactifying this metric on $T^6$ we arrive at
the string in $D=5$ which has a constant dilaton and the following
metric:

\begin{equation}
\d s_{(5)}^2 = (f_1 f_2 f_3)^{-1/3} 
    \left[ - f_4^{-1} h \d t^2 + 
f_4 \left( \d y - {Q\over r+r_4}\d t\right)^2 \right] + 
   (f_1 f_2 f_3)^{2/3} \left(h^{-1} \d r^2 + r^2 \d \Omega^2 \right)  
                                                    \label{5DimMetric}
\end{equation}

\noindent
where

\begin{equation}
f_i = 1 + \df{r_i}{r} \ , \qquad h = 1 - \df{r_0}{r}\ .     
\end{equation}

\noindent 
It is useful to define a hyperbolic angle $\sigma$ such that

\begin{equation}
r_4 = r_0 \sinh^2 \sigma\ , \qquad
Q= r_0 \sinh \sigma~ \cosh \sigma \ .
\end{equation}

\noindent
The quantities $r_4$ and $Q$ are comparable to the horizon radius
$r_0$. Therefore $\sigma$ is of order one. 
Dimensional reduction on $S^1$ from $D=5$ to $D=4$ gives a
black hole with four $U(1)$ charges described by the following metric
and Kaluza-Klein gauge potential $A_0$:

\begin{eqnarray}
\d s_{(4)}^2 &=& -f^{-1/2}h \d t^2 + 
                  f^{1/2} \left(h^{-1} \d r^2 + r^2 \d \Omega^2 \right)
\nonumber \\
f &=& \prod_{i=1}^4 \left( 1 + {r_i\over r} \right)  \nonumber \\
A_0 &=& Q/(r_4+r) \ .                          \label{4DimMetric}
\end{eqnarray}

\noindent
For the region of parameters \equno{newchoice}, the Bekenstein-Hawking
entropy of this black hole is \cite{ct1,kt1}

\begin{equation}
S_{\rm BH} = {2\pi\over \kappa_4^2} A_{\rm h}=
{8\pi\over \kappa_4^2} \sqrt{r_1 r_2 r_3 r_0} \cosh \sigma \ .
\label{bekh}\end{equation}

\noindent
Calculation of the Hawking temperature gives
\cite{ct1,kt1}

\begin{equation}
{1\over T_H} = 4 \pi \sqrt{r_1 r_2 r_3\over r_0} \cosh \sigma \ .
\end{equation}

\noindent
We may write ${2\over T_H}={1\over T_L}+{1\over T_R}$, where

\begin{equation}
{1\over T_L}= 4 \pi \sqrt{r_1 r_2 r_3\over r_0} e^{-\sigma}\ ,
\qquad
{1\over T_R}= 4 \pi \sqrt{r_1 r_2 r_3\over r_0} e^{\sigma} \ .  \label{TLTR}
\end{equation}

\noindent
In the effective string model $T_L$ and $T_R$ have the meaning
of left and right moving temperatures. Indeed, 
it may be shown that the entropy
\equno{bekh} equals $\pi \Leff (T_L+ T_R)$, which is the entropy
of a string of length $\Leff$, carrying massless modes of central
charge $c=6$.

We are interested in studying the propagation of scalars carrying 
Kaluza-Klein charge in the background of this charged black hole.
This is best viewed as propagation of scalars in the background of the
$D=5$ string, with the role of the charge being played by the momentum
along the string, $k_4$. Thus, we substitute the
ansatz
$\phi(t,y,r) = \e^{-\i \omega t} \e^{-\i k_4 y} R(r)$ 
into the $D=5$ scalar equation

\begin{equation}
\df{1}{\sqrt{-g^{(5)}}} \partial_M \left (\sqrt{-g^{(5)}}
  g_{(5)}^{MN} \partial_N \phi \right )= 0 \ .
\end{equation}

\noindent
The resulting radial equation is

\begin{equation}
\prod_{i=1}^3 \left (1+{r_i\over r} \right )
\left [ \omega^2 - k_4^2 + (\omega \sinh \sigma - k_4 \cosh\sigma)^2 
{r_0\over r} \right ] R+ 
{h\over r^2}{\d \over \d r} h r^2 {\d R \over \d r}=0 \ .
\label{charged}\end{equation}

\noindent
This is very similar to the radial equation found for the $D=5$
black hole \cite{mast}. It is again possible to define new variables

\begin{equation}
\omega'^2= \omega^2- k_4^2 \ , \qquad 
e^{\pm \sigma'}=e^{\pm \sigma} {(\omega \mp k_4)\over \omega'}\ ,  
\label{Primed} \end{equation}

\noindent
such that \equno{charged} becomes
\begin{equation}
\omega'^2\left (1+{r_0\sinh^2\sigma'\over r}\right )
\prod_{i=1}^3 \left (1+{r_i\over r}\right ) R
+ {h\over r^2}{\d \over \d r} h r^2 {\d R \over \d r}=0 \ ,
\label{neutral}\end{equation}

\noindent
which describes propagation of a neutral particle of energy
$\omega'$ near a black hole with the hyperbolic angle parameter
$\sigma$ replaced by $\sigma'$.
The absorption cross section may be calculated using the matching
method in a manner similar to \cite{mast}.

Rather than run through the beautiful matching calculation of \cite{mast}
step by step, we will give only a summary designed to
provide some continuity with the analysis of previous papers \cite{dmw,dm,us}.
We will restrict our attention to a neutral scalar of energy $\omega$
propagating in the black hole background described by \equno{4DimMetric},
that is to say by unprimed variables.  At the end we will recover the
general charged case using the primed variables defined in \equno{Primed}.

A variety of radial coordinates are useful in different contexts.  The ones
we will employ are related as follows:

\begin{equation}
z = 1 - {r_0 \over r} = \e^{-r_0/u} \ .                \label{RadVariables}
\end{equation}

\noindent
This equation can be taken as the definition of $z$ and $u$.  The 
coordinate $u$ is useful because it brings the equation \equno{neutral} 
into the simple form

\begin{equation}
\left( \df{1}{u^2} \df{\d}{\d u} u^2 \df{\d}{\d u} + 
  \df{r^4}{u^4} f \omega^2 \right) R = 0 \ ,            \label{Exact}
\end{equation}

\noindent
and because the radial flux per unit solid angle is just 

\begin{equation}
{\cal F} = \df{1}{2 \i} \left( R^* u^2 \df{\d}{\d u} R - {\rm c.c.} 
   \right) \ .
                                                       \label{RadFlux}
\end{equation}
\noindent
The matching calculations that produced the results of \cite{dmw,dm,us} 
amount essentially to the following: we simplify \equno{Exact} in the 
regions near the horizon ({\bf I}) and far from the black hole ({\bf III})
by keeping only the leading term in a small $u$ or large $u$ expansion of
the term $f r^4 / u^4$.  Modulo some technical assumptions, 
the near and far 
solutions can then be matched directly onto one another.  In the present
instance, one obtains

\begin{eqnarray}
R_{\bf I} &=& A \e^{\i \sqrt{P} / u} \qquad {\rm where} \qquad 
  P = \omega^2 \prod_{i = 1}^4 (r_i + r_0)              \label{RI}  \\
R_{\bf III} &=& \alpha \df{\sin \omega u}{\omega u} - 
              \beta \df{\cos \omega u}{\omega u}        \label{RIII}
\end{eqnarray}

\noindent
and these can be matched by making expansions of $R_{\bf I}$ and $R_{\bf
III}$ for large and small $u$ respectively.  The S-matrix element $S_0$ for 
reflection of the $s$-wave can be read off from comparison of \equno{RIII}
to the standard form

\begin{equation}
R \sim \df{S_0 \e^{\i \omega u} - \e^{-\i \omega u}}{\omega u} \ ,
\end{equation}

\noindent
but a simpler method employed by \cite{mast} is to note that the absorption
probability is the ratio of two fluxes:

\begin{equation}
1 - |S_0|^2 = {{\cal F}_{\rm absorbed} \over {\cal F}_{\rm incoming}} \ .
\end{equation}

\noindent
${\cal F}_{\rm incoming}$ is computed by applying \equno{RadFlux} to 
the incoming wave $(\alpha + \i \beta) \i \e^{-\i \omega u} / (2 \omega u)$
at infinity.  We consider
only perturbative scattering processes where
$|\beta| \ll |\alpha|$, so the leading order result depends only on 
$\alpha$.  Similarly, ${\cal F}_{\rm absorbed}$ is computed by applying
\equno{RadFlux} to $R_{\bf I} = A \e^{\i \sqrt{P} / u}$ at the horizon.  The 
result for the absorption probability is 

\begin{equation}
1 - |S_0|^2 = 4 \omega \sqrt{P} \df{|A|^2}{|\alpha|^2} \ .
\end{equation}

\noindent
The optical theorem converts this probability into a cross section:

\begin{equation}
\sigma_{\rm abs} = \df{\pi}{\omega^2} \left( 1 - |S_0|^2 \right) 
  = A_{\rm h} |E|^{-2}
\end{equation}

\noindent
where we have defined the greybody factor $E = \alpha / A$.  

The simplicity of the ratio of fluxes method outlined above is that $E$ 
can be obtained by matching the limiting value of $R_{\bf I}$ at 
large $u$ with the limiting value of $R_{\bf III}$ at small $u$.  Of 
course, one loses all phase information for the reflected wave in this 
approach, and one must still check that a full matching is possible.  

Matching \equno{RI} and \equno{RIII} in the manner described yields
$E = 1$ and hence $\sigma_{\rm abs} = A_{\rm h}$. 
This most naive matching scheme fails when one gives up the condition
$r_0 \ll r_n$ because there are neglected terms in the near region 
equation which are of order $r_0 / r_n$.  Following the approach of
\cite{mast}, we retain all such terms and exclude only terms of 
order $r_0 / r_i$ for $i = 1,2,3$.  The effect on the infalling 
solution $R_{\bf I}$ is to introduce a modulating factor $F(u)$:  
instead of \equno{RI} we now have

\begin{equation}
R_{\bf I} = A \e^{\i \sqrt{P} / u} F(u) \ .                \label{RINew}
\end{equation}

\noindent
The boundary conditions $F(u) = 1$ and $u^2 F'(u) = 0$ are imposed at 
$u = 0$, so ${\cal F}_{\rm absorbed}$ is the
same as one would calculate from the bare exponential \equno{RI}.  $E$
is read off as the limiting value of $F(u)$ for large $u$.  
Starting
from \equno{Exact} and neglecting $r_0 / r_i$ for
$i = 1,2,3$, we find that
$F(u)$ satisfies a hypergeometric equation in the
variable $z$:

\begin{equation}
\left[ z (1-z) \df{\d^2}{\d z^2} + 
  (1-z) \left( 1 - \i (a+b) \right) \df{\d}{\d z}  + ab \right] F = 0 \ ,
                                                           \label{HyperGeom}
\end{equation}

\noindent
where the parameters $a$ and $b$ are given by 

\begin{equation}
a = \df{\omega}{4 \pi T_R} \qquad  b = \df{\omega}{4 \pi T_L} \ ,
                                                           \label{abParams}
\end{equation}

\noindent
and the temperatures $T_R$ and $T_L$ are given by \equno{TLTR}.
Interestingly, 
\equno{HyperGeom} and \equno{abParams} are identical to the 
inner region equations for the $D=5$ black hole \cite{mast}, although
$z$, $T_L$ and $T_R$ are defined differently.

Once it is established that $F(z)$ is a hypergeometric function, $E$ 
can be read off directly from the asymptotics of $F(z)$ near $z = 1$
\cite{mast}:

\begin{equation}
E = \df{\Gamma(1 - \i a - \i b)}{\Gamma(1 - \i a) \Gamma(1 - \i b)} \ .
                                                       \label{EGamma}
\end{equation}

\noindent
The formula for the absorption cross section then becomes

\begin{equation}
\sigma_{\rm abs} = A_{\rm h} \df{\omega }{2(T_L + T_R)}
  {\e^{\omega\over T_H} - 1\over \left (\e^{\omega\over 2 T_L} - 1\right )
  \left (\e^{\omega\over 2 T_R} - 1 \right ) } \ .        \label{SigmaUniv}
\end{equation}

\noindent
The cross section for the five-dimensional case can also be written in 
precisely this form. As $\omega\rightarrow 0$, 
$\sigma_{\rm abs} \rightarrow A_{\rm h}$
in accord with the general result of \cite{dgm}.
Equation \equno{SigmaUniv} contains even more universal
information: it captures the behavior of the cross-section as $\omega$
is increased to values that are comparable with the temperatures.

Restoring primes, we obtain for the charged case

\begin{equation}
\sigma_{abs} =4\pi^2 r_1 r_2 r_3 \omega'
{\e^{\omega'\over T_H'} - 1\over \left (\e^{\omega'\over 2 T_L'} - 1\right )
\left (\e^{\omega'\over 2 T_R'} - 1 \right ) } \ .
\end{equation}

\noindent
To write this expression in terms of physical quantities we use the formulae
\cite{mast}

\begin{equation}
{\omega'\over T_L'} = {\omega + k_4\over T_L}\ ,
\quad {\omega'\over T_R'} = {\omega - k_4\over T_R}\ ,
\quad {\omega'\over T_H'} = {\omega -\phi k_4\over T_H}\ ,
\end{equation}

\noindent
where $\phi=A_0 (r_0)= \tanh \sigma$ is the $U(1)$ potential on the
horizon. Thus, for a particle of energy $\omega$ and charge $k_4$,
the absorption cross section is

\begin{equation}
\sigma_{abs} = 4\pi^2 r_1 r_2 r_3 \sqrt{ \omega^2 - k_4^2}
{\e^{\omega -\phi k_4\over T_H} - 1\over 
\left (\e^{\omega + k_4\over 2 T_L} - 1\right )
\left (\e^{\omega - k_4\over 2 T_R} - 1 \right ) } \ .
\end{equation}

\noindent
{}From this we may obtain the differential Hawking emission rate,

\begin{eqnarray}
\Gamma (\vec{k}) \df{\d^3 k}{(2 \pi)^3}
&=& {\sqrt{ \omega^2 - k_4^2}\over \omega} \sigma_{abs}
{1\over \e^{\omega -\phi k_4\over T_H} - 1} {d^3 k\over (2\pi)^3}  \nonumber \\
&=& 4 \pi^2 r_1 r_2 r_3 {\omega^2 - k_4^2\over \omega}
{1\over
\left (\e^{\omega + k_4\over 2 T_L} - 1\right )
\left (\e^{\omega - k_4\over 2 T_R} - 1 \right ) }
{d^3 k\over (2\pi)^3} \ .
\label{semrate}\end{eqnarray}

Remarkably, this formula is in precise agreement with the 
effective string model. Indeed, in \cite{us} it was
shown that this model predicts the leading order Hawking rate

\begin{equation}
\Gamma (\vec{k}) \df{\d^3 k}{(2 \pi)^3} =
  \df{\kappa_4^2 \Leff}{4} \df{\omega^2-k_4^2}{\omega}
{1\over
\left (\e^{\omega + k_4\over 2 T_L} - 1\right )
\left (\e^{\omega - k_4\over 2 T_R} - 1 \right ) }
  \df{\d^3 k}{(2 \pi)^3} \ ,                            \label{newrate}
\label{stringrate}\end{equation}

\noindent
where $\Leff= n_1 n_2 n_3 L_1$ and $L_1$ is the length of the circle
around which the string is wound.
This rate is due to a left-moving boson and
a right-moving boson on the string producing an outgoing scalar, hence the 
presence of the two Bose-Einstein distribution factors.
The string effective lagrangian contains no cubic term coupling a
left-moving fermion and
a right-moving fermion to a scalar in the bulk; therefore, there
is no additive contribution containing two Fermi-Dirac distributions.

The radii $r_i$ (approximately equal to
the charges $Q_i$) are related to the numbers of 5-branes \cite{kt}:

\begin{equation}
r_1 = \df{n_1}{L_6 L_7} \left( \df{\kappa_{11}}{4 \pi} \right)^{2/3} \qquad 
r_2 = \df{n_1}{L_4 L_5} \left( \df{\kappa_{11}}{4 \pi} \right)^{2/3} \qquad
r_3 = \df{n_1}{L_2 L_3} \left( \df{\kappa_{11}}{4 \pi} \right)^{2/3} \ ,
\end{equation}

\noindent
where $L_i$ is the range of the coordinate $y_i$.
Thus, we have

\begin{equation}
\df{\kappa_4^2 \Leff}{4}=  4 \pi^2 r_1 r_2 r_3 
\label{chargerel}\end{equation}

\noindent
where we have used $\kappa_4^2= \kappa_{11}^2/\prod_{i=1}^7 L_i$.
Equation \equno{chargerel} establishes exact equality between
\equno{stringrate}, the Hawking rate for charged particles
calculated in the effective
string model, and \equno{semrate},
the corresponding rate calculated in semiclassical gravity.

In closing we would like to reflect on the significance of our 
result.  Because the M-theory description of
four-dimensional black holes is not as developed as
the D-brane description of
five-dimensional ones, it seems to us very encouraging to find that the
greybody factors in the four-dimensional case confirm the underlying 
effective string description which was previously used in
\cite{kt,us,kt1}. 
We regard \equno{semrate} as ``new data'' from
semiclassical relativity, obtained in the region
\equno{newchoice} of parameter space,
that supports the claim that black holes in four
dimensions admit an effective string description.

\section*{Acknowledgements}

We are grateful to 
A.~Tseytlin for discussions.  The work of
I.R.~Klebanov was supported in part by DOE grant DE-FG02-91ER40671,
the NSF Presidential Young Investigator Award PHY-9157482, and the
James S.{} McDonnell Foundation grant No.{} 91-48.  S.S.~Gubser 
thanks the Hertz Foundation for its support.



\end{document}